\documentclass{svmult} 

\usepackage{graphicx}        
\usepackage[bottom]{footmisc}

\begin{document}

\title*{The Schr\"odinger Equation, Reversibility and the Grover Algorithm}
\author{Miguel A. Martin-Delgado}
\institute{Departamento de
F\'{\i}sica Te\'orica I, Universidad Complutense. 28040 Madrid, Spain.
\texttt{mardel@miranda.fis.ucm.es}
}

\maketitle

Several aspects of the time-dependent Schr\"odinger equation are
discussed in the context of Quantum Information Theory.

\section{Time Dependent Schr\"odinger Equation and Quantum Computation}
\label{sec:1}

In Quantum Information Theory (QIT), the time dependent Schr\"odinger
equation \cite{gp90} acquires an informational content in 
addition to its physical
content that dictates the time evolution of a quantum system.
The functioning of a Quantum Computer (QC) \cite{rmp} parallels the unitary
evolution of a quantum system. Each step of the quantum evolution
corresponds to a typical operation on a QC:

\noindent i/  A  QC starts at some initial state 
\begin{equation}
|\Psi(t_{\rm i})\rangle := |\Psi_0\rangle.
\label{a1}
\end{equation}
The preparation of the initial state corresponds to the input data
in a computer program. The state $|\Psi(t_{\rm i})\rangle$ belongs 
to a Hilbert space
of many-particle states ${\cal H}^{\otimes n}$ that is a tensor product of 
$n$ single-particle states. When each single Hilbert space ${\cal H}$
is spanned by a two-dimensional basis, the information is quantumly 
encoded in a qubit. A quantum register is a tensor product of qubit
states. Quantum states can be used to encode information.
A manageable configuration  for the initial state is a factorized
tensor product of qubits (unentangled state)
\begin{equation}
|\Psi(t_{\rm i})\rangle = |x_1\rangle \otimes |x_2\rangle \otimes
\ldots \otimes |x_n\rangle \in {\cal H}^{\otimes n}, \; x_i=0,1 \;\forall i.
\label{a1b}
\end{equation}

\noindent ii/ A QC  evolves in  time during an interval  $t$ allowing us to
process the information previously encoded 
\begin{equation}
|\Psi(t)\rangle = U(t,t_{\rm i})|\Psi(t_{\rm i})\rangle.
\label{a2}
\end{equation}
where the unitary evolution operator $U(t,t_{\rm i})$ satisfies
the time-dependent Schr\"odinger equation, either in differential
form
\begin{equation}
{\rm i}\hbar \frac{d}{dt}U(t,t_{\rm i}) = H(t)U(t,t_{\rm i}),
\label{a3}
\end{equation}
or as an equivalent integral equation
\begin{equation}
U(t,t_{\rm i}) = I - \frac{\rm i}{\hbar} 
\int_{t_{\rm i}}^t H(t')U(t',t_{\rm i}) dt'.
\label{a4}
\end{equation}
The integration of the evolution equation yields
\begin{equation}
U(t,t_{\rm i}) = {\rm T}\left[{\rm e}^{-\frac{{\rm i}}{\hbar}
\int_{t_{\rm i}}^t H(t') dt'}\right],
\label{a4b}
\end{equation}
where T is the temporal ordering operator introduced by Dyson.
When the Hamiltonian is time independent as in conservative systems, 
the integrated form of the evolution operator can be further simplified as
\begin{equation}
U(t,t_{\rm i}) = {\rm e}^{-\frac{{\rm i}}{\hbar}(t-t_{\rm i})H}.
\label{a5}
\end{equation}
The evolution operator of Quantum Mechanics corresponds to a quantum logic
gate in a QC. 
Thus, a quantum logic gate is a unitary operator acting on the states of a
certain set of qubits (quantum register). It is represented by a 
$2^n\times 2^n$ matrix in 
the unitary group ${\rm U}(2^n)$. Hence, quantum computation is a class
of reversible computation. The first experimental proposal for constructing
quantum logic gates with ion traps was by Cirac and Zoller 
\cite{cz1},\cite{cz2}.

\noindent iii/ Quantum programming corresponds to harnessing the reversibility
inherent to quantum computation with the purpose of producing a useful task.
A task corresponds to the evaluation of a certain Boolean function
$f: \{0,1\}^m \to \{ 0,1\}$.
Reversibility  requires to split the
quantum register storing an initial state $|\Psi_{\rm i}\rangle$
into two parts: the {\em source register} and the 
{\em target register}, namely,
\begin{equation}
|\Psi_0\rangle := |\Psi_{\rm s}\rangle \otimes |\Psi_{\rm
t}\rangle.
\label{a6}
\end{equation}
To implement a Boolean function  $f$ in a QC we need the
action of a unitary gate $U_f$ on the registers as follows
\begin{equation}
U_f|x_1x_2\ldots x_m\rangle_{\rm s} |x_{m+1}\rangle_{\rm t}=
|x_1x_2\ldots x_m\rangle_{\rm s} |x_{m+1}\oplus
f(x_1,x_2,\ldots,x_m)\rangle_{\rm t}.
\label{a7}
\end{equation}
This action guarantees that whichever the $f$ is, the function evaluation is
implemented reversibly \cite{functionevaluation}.

\noindent iv/ The final state $|\Psi_{\rm f}\rangle$ is the 
outcome of a computation in a QC. It is a superposition of all
states in a computational basis $\{ |x_1,\ldots,x_n\rangle\}$
\begin{equation}
|\Psi_{\rm f}\rangle = U(t_{\rm f},t_{\rm i})|\Psi(t_{\rm i})\rangle
= \sum_{x_1,\ldots,x_n=0,1} C_{x_1,\ldots,x_n}|x_1,\ldots,x_n\rangle,
\label{a8}
\end{equation}
and will be an entangled state in contrast to the factorized state that
we started with (\ref{a1b}).
Entanglement is the result of combining the superposition principle
of Quantum Mechanics with the tensor product of single-particle systems.
This capability of storing many classical registers $(x_1,\ldots,x_n)$ at the
same time (\ref{a8}) is called quantum parallelism. It is the source of
the potentially superior efficiency of a QC with respect to a classical
computer.

\noindent v/ The measurement in Quantum Mechanics is the read out of the
QC to retrieve the desired output. Associated to the computational basis
there is a set of orthonormal operators 
$\Pi_{(x_1,\ldots,x_n)}=|x_1,\ldots,x_n\rangle\langle x_1,\ldots,x_n|$
such that the result of a measurement on the final state is
\begin{equation}
|\Psi_{\rm f}\rangle \langle \Psi_{\rm f}| \longrightarrow
\frac{\Pi_{(x_1,\ldots,x_n)} |\Psi_{\rm f}\rangle \langle \Psi_{\rm f}| \Pi_{(x_1,\ldots,x_n)}}{{\rm Tr}\left[ \Pi_{(x_1,\ldots,x_n)} |\Psi_{\rm f}\rangle \langle \Psi_{\rm f}| \right]} = \Pi_{(x_1,\ldots,x_n)},
\label{a9}
\end{equation}
with probability
\begin{equation}
{\cal P}(x_1,\ldots,x_n) = {\rm Tr}\left[ \Pi_{(x_1,\ldots,x_n)} |\Psi_{\rm f}\rangle \langle \Psi_{\rm f}| \right] =
|C_{x_1,\ldots,x_n}|^2.
\label{a10}
\end{equation}
Thus, the probabilistic nature of Quantum Mechanics implies that a quantum
computation is probabilistic: we will find the desired output within a 
certain probability (\ref{a10}). The aim of a good quantum programmer is
to make this probability as close as possible to 1. This means a 
pattern of constructive interference of amplitudes towards the 
desired output amplitude.
Examples of good quantum programming are the Shor algorithm \cite{shor}, where
the task is to factor out a large integer, and the Grover
algorithm \cite{grover2}, 
where the task is to search for an item in a disordered database.
 
This is the  time evolution of a quantum system and it encompasses all 
the principles of Quantum Mechanics, which in turn can aslo be given an 
interpretation in QIT
\cite{principlesQIT}.

The basic problems that a QC can address are the same than a classical
computer:

\begin{itemize}

\item {\em Computability}:  what problems the machines can do and
cannot do.

\item {\em Complexity}:  scaling of space/time with the size of the
problems that can be solved.

\item {\em Universality}:  whether one machine can simulate all others,
so that we do not need to construct an special-purpose machine for each
problem to be solved.

\end{itemize}

\section{Reversibility on the Lattice}
\label{sec:2}

Reversibility means that we can compute exactly backwards 
by saving only the final conditions and the number of steps. 
It is energetically
economical compared to current standard computers that work irreversibly
\cite{reversibility}.

A QC works reversibly as a consequence of the unitary of the 
evolution operator (\ref{a5}). It is also a machine that operates with
finite means (Turing machine): 
to simulate the evolution of a quantum system we must
discretize both time and space. This is done by setting the Schr\"odinger
equation on a space-time lattice. This discretization was implicit in the
qubit decomposition of the initial state in (\ref{a1b}).
From a more fundamental point of view, many theoretical physicists, including
Feynman \cite{hillis}, have wondered about what is the underlying structure
of the universe beyond the subatomic level and have proposed that it might
have a discrete structure. If this were the case, a natural question is
how to formulate reversible evolutions in a discretized form.

The finite difference method  is a natural scheme to discretize the  
Schr\"odin\-ger equation: substitute derivatives of functions by finite
differences, which are approximations to a certain degree in a 
Taylor expansion. There is some freedom in this transition
from a differencial equation to its finite difference form. It turns out that
some of them are not exactly reversible while some of them are. 
The meaning here of the word exactly  is explained below in the context
of computer simulations with finite precission.

The key point is whether we use a naive difference formula for the
derivative of a given function $F(z)$ as
\begin{equation}
\frac{d}{dz}F(z) = \frac{1}{\epsilon}[F(z_{i+1})-F(z_i)] + O(\epsilon).
\label{b1}
\end{equation}
This is a non-centered formula. Another possibility is to use centered
formulas for derivatives of any degree
\begin{equation}
\frac{d}{dz}F(z) = \frac{1}{2\epsilon}[F(z_{i+1})-F(z_{i-1})] + O(\epsilon^2),
\label{b2}
\end{equation}
\begin{equation}
\frac{d^2}{dz^2}F(z) = \frac{1}{\epsilon^2}[F(z_{i+1})-2F(z_i)+F(z_{i-1})]+ 
O(\epsilon^2).
\label{b3}
\end{equation}

Let us consider the evolution of a single particle in one dimension
given by the Schr\"odinger equation in coordinate representation
\begin{equation}
{\rm i}\frac{\partial}{\partial t}\Psi(x,t) = - \frac{\partial^2}{\partial x^2}\Psi(x,t) + V(x,t) \Psi(x,t),
\label{b4}
\end{equation}
where $V(x,t)$ is a generic potential and we assume units $\hbar=2m=1$. 
To discretize this equation,
let $a$ be the lattice spacing so that the discrete coordinates are
$x_m=ma, m\in Z$ and $\tau$ the time step such that time coordinates
are $t_n=n\tau, n\in Z$. The discretized wave function is 
$\Psi_{m,n}:=\Psi(x_m,t_n)$ and similarly for the potential
$V_{m,n}:=V(x_m,t_n)$.

\subsubsection{Asymmetric Difference Scheme}

This is a (centered space, non-centered time) difference scheme that
results from the substitution of (\ref{b1}) for the time derivative
and (\ref{b3}) for the spatial derivative 
\begin{equation}
\Psi_{m,n+1} = \Psi_{m,n} + {\rm i}\left[
\frac{\tau}{a^2}
\left( \Psi_{m+1,n}-2\Psi_{m,n}+\Psi_{m-1,n} \right) -
\tau V_{m,n}\Psi_{m,n}\right].
\label{b5}
\end{equation}
This is a first order difference equation in the time index $n$ like
its continuum version. The initial value problem is determined by
giving the values $\Psi_{m,0}$ along the one-dimensional spatial grid.
However, this equation (\ref{b5}) is not an exactly reversible discrete update rule because of roundoff and truncation errors.
The matrix form of (\ref{b5}) is, for periodic boundary conditions in a lattice
with $M$ sites,
\begin{equation}
{\vec \Psi}_{n+1} = 
\left(\begin{array}{cccccc}
1-{\rm i}V_{1,n}\tau - 2{\rm i}\epsilon & 
{\rm i}\epsilon & 
0 & 
0 &
\ldots & 
{\rm i}\epsilon \\
{\rm i}\epsilon & 
1-{\rm i}V_{2,n}\tau - 2{\rm i}\epsilon &
{\rm i}\epsilon & 
0 &
\ldots & 
0 \\
\vdots &
\vdots &
\vdots &
\vdots &
\ddots &
\vdots \\
{\rm i}\epsilon & 
0 &
0 &
0 &
\ldots &
1-{\rm i}V_{M,n}\tau - 2{\rm i}\epsilon
\end{array} \right)
\vec{\Psi}_{n},
\label{b6}
\end{equation}
where $\epsilon:=\frac{\tau}{a^2}$ and $\vec{\Psi}_n$ is a column vector
of dimension $M$ storing the values of the wave function at time step $n$.
This matrix is unitary only in the limit of infinitesimal $\epsilon$ and
$\tau$: the norm of each column vector is $1+O(\tau^2)+O(\epsilon^2)$,
and the scalar product of two vectors is $O(\tau^2)+O(\epsilon^2)$.
Were it be exactly unitary, its inversion would not need 
multiplication/division operations since it would be its Hermitian matrix: 
only simple arithmetic operations
would do the job.

\subsubsection{Symmetric Difference Scheme}

This is a (centered space, centered time) difference scheme that
results from the substitution of (\ref{b2}) for the time derivative
and (\ref{b3}) for the spatial derivative 
\begin{equation}
\Psi_{m,n+1} = \Psi_{m,n-1} + {\rm i}\left[
2\frac{\tau}{a^2}
\left( \Psi_{m+1,n}-2\Psi_{m,n}+\Psi_{m-1,n} \right) -
2\tau V_{m,n}\Psi_{m,n}\right].
\label{b7}
\end{equation}
This scheme was discovered by Fredkin and Barton in 1975 \cite{fredkin} 
in their quest for exact reversibility of discretized Schr\"odinger equation
following discussions with Feynman \cite{fredkin}, \cite{frank}, \cite{motter}.
This discretized update rule
have very interesting properties:

\begin{description}

\item[ i/] It is a second order difference equation:
the initial value problem is defined by providing the values of the
wave function at two earlier time steps. This is in sharp contrast
with the original continuous differential equation.

\item[ii/] It is exactly reversible after $n$ steps of computation.

\item[iii/] It is exactly reversible even if the RHS cannot be
computed exactly.

\item[iv/]  It is exactly reversible even if $\Psi_{m,n}$ has
finite precision values.

\item[v/] It is a complex equation. However, the real part of 
$\vec{\Psi}_{n+1}$ depends only on the real part of $\vec{\Psi}_{n-1}$ and
the imaginary part of $\vec{\Psi}_{n}$.
Likewise, the imaginary part of $\vec{\Psi}_{n+1}$ depends only on the 
imaginary part of $\vec{\Psi}_{n-1}$ and
the real part of $\vec{\Psi}_{n}$.

\end{description}

The exact reversibility expressed in properties ii/,iii/ and iv/ is related
to property i/ and follows from the backwards recurrence relation:
\begin{equation}
\Psi_{m,n-1}  =  \Psi_{m,n+1} - {\rm i}\left[
2\frac{\tau}{a^2}
\left( \Psi_{m+1,n}-2\Psi_{m,n}+\Psi_{m-1,n} \right) -
2\tau V_{m,n}\Psi_{m,n}\right].
\label{b8}
\end{equation}
Property v/ is also useful to see the exact reversibility more explicitly:
we can select the real parts of the wave function at all even time
steps $n=2l$ and the imaginary parts at all odd time steps $n=2l+1$.
Thus, v/ guarantees that the evolution under (\ref{b7}) respects this structure
and the rest of components of $\vec{\Psi}_{2l}, \vec{\Psi}_{2l+1}$ 
can be ignored self-consistently.

As a warmup, let us consider a typical update rule for a set of variables
$\{ X_{2l}, Y_{2l+1} \}_{l=0}^{\infty}$, ordered according to the time
evolution of the real and imaginary parts described above: 
$X_0,Y_1,X_2,Y_3,X_4, Y_5,X_6,\ldots$. The initial data is $X_0,Y_1$.
Let us assume that the update rule has the following form:
\begin{equation}
\left\{ 
\begin{array}{ccc}
X_{2l+2} & := & X_{2l} + F(Y_{2l+1}),\\ 
Y_{2l+3} & := & Y_{2l+1} + F(X_{2l}), \\ 
\end{array}
\right.
\label{b9}
\end{equation}
where $F$ is an arbitrary function. This is an example of an exactly 
reversible recurrence relation: the backwards update rule follows inmediately,
\begin{equation}
\left\{ 
\begin{array}{ccc}
Y_{2l+1} & = & Y_{2l+3} - F(X_{2l}),\\ 
X_{2l} & = & X_{2l+2} - F(Y_{2l+1}). \\ 
\end{array}
\right.
\label{b10}
\end{equation}
Notice that in doing this reverse operation there is no need for 
multiplication/division
operations, instead the simple arithmetic of addition/substraction 
can do the job.
This fact is important when implementing the exact reversibility on a 
computer.

Let us now turn to the evolution difference equations for the real and
imaginary parts of the wave function:
\begin{equation}
\Psi_{m,n} := R_{m,n}  + {\rm i} I_{m,n}.
\label{b11}
\end{equation}
From (\ref{b7}), they are given by
\begin{equation}
\left\{ 
\begin{array}{ccc}
 R_{m,n+1}& = & R_{m,n-1} -\left[
2\epsilon \left( I_{m+1,n}-2I_{m,n}+I_{m-1,n} \right) -
2\tau V_{m}I_{m,n}\right],\\ 
I_{m,n+1} & = & I_{m,n-1} +\left[
2\epsilon \left( R_{m+1,n}-2R_{m,n}+R_{m-1,n} \right) -
2\tau V_{m}R_{m,n}\right]. \\ 
\end{array}
\right.
\label{b12}
\end{equation}
For simplicity we have assumed that the potential is time-idependent
$V_{m,n}:=V_m, \forall n.$
Thus, setting $n=2l+1$ and $n=2l+2$ in the first and second equations of 
(\ref{b12}) respectively, we can discard the real parts at all odd time
steps and the imaginary parts at all even time steps, yielding
\begin{equation}
\left\{ 
\begin{array}{ccc}
 R_{m,2l+2}& = & R_{m,2l} -\left[
2\epsilon \left( I_{m+1,2l+1}-2I_{m,2l+1}+I_{m-1,2l+1} \right) -
2\tau V_{m}I_{m,2l+1}\right],\\ 
I_{m,2l+3} & = & I_{m,2l+1} +\left[
2\epsilon \left( R_{m+1,2l+2}-2R_{m,2l+2}+R_{m-1,2l+2} \right) -
2\tau V_{m}R_{m,2l+2}\right]. \\ 
\end{array}
\right.
\label{b13}
\end{equation}
They have precisely the same form as (\ref{b9}) 
\begin{equation}
\left\{ 
\begin{array}{ccc}
R_{m,2l+2}& = & R_{m,2l} - F_m(I_{m,2l+1}),\\ 
I_{m,2l+3} & = & I_{m,2l+1} + F_m(R_{m,2l}), \\ 
\end{array}
\right.
\label{b14}
\end{equation}
with $F_m$ given by
\begin{equation}
F_m(C_{m,n}):= 2\epsilon \left( C_{m+1,n}-2C_{m,n}+C_{m-1,n} \right) -
2\tau V_{m}C_{m,n}.
\label{b15}
\end{equation}
Therefore, we have an exactly reversible update rule for the real and
imaginary parts of the wave function $\vec{\Psi}_{n}$ at successive time steps.
This does not produce the values of $\vec{\Psi}_{n}$ at arbitrary times, but
assuming that the change in $\vec{\Psi}_{n}$ is smooth when $\tau$ is small 
enough, then we can define the complex value of the wave function at 
any time as
\begin{equation}
\Psi_{m,l} := R_{m,2l} + {\rm i} I_{m,2l+1}.
\label{b16}
\end{equation}

A more compact form of the update recursion relations are obtained if 
we arrange the spatial components of the wave function in a vector array:
\begin{equation}
\vec{R}_{2l} := (R_{m,2l}), \; \vec{I}_{2l+1}:=(I_{m,2l+1}).
\label{b17}
\end{equation}
Moreover, we may want to implement the evolution equations directly
in a computer with only integer arithmetics and resorting to 
only summation/substraction operations. The update rules (\ref{b14}) 
are precisely well suited for this type of implementations using for 
instance the C/C++ languaje operations \texttt{+=} and \texttt{-=}. 
With these provisos,
the computer-like form of the evolution difference equations is
\begin{equation}
\left\{
\begin{array}{ccc}
\vec{R} \; \; &-=& \lfloor \vec{F}(\vec{I})\rfloor,\\
\vec{I} \; \; &+=& \lfloor \vec{F}(\vec{R})\rfloor .
\end{array}
\right.
\label{b18}
\end{equation}

The total probability  $P(t):=||\Psi(x,t)||^2$ for finding the particle
is a conserved quantity in the 
evolution of the quantum system (\ref{b4}). This quantity admits a 
discretized implementation like
\begin{equation}
P_{l} := \vec{R}_{2l}^2 + \vec{I}_{2l+1}^2.
\label{b19}
\end{equation}
However, it is not exactly reversible for the discretized time
evolution (\ref{b14}). Feynman found \cite{fredkin}
the following invariant
\begin{equation}
P_{l}^{({\rm r})} := \vec{R}_{2l}^2 - \vec{I}_{2l+1}\cdot \vec{I}_{2l-1},
\label{b20}
\end{equation}
that is exactly reversible under the evolution (\ref{b14}).

The stability of the exactly reversible Schr\"odinger equation
is analysed in appendix \ref{sec:apA}.

So far we have considered the reversible vs. non-reversible implementation 
of the time-dependent Schr\"odinger equation in a classical computer.
We can also implement it in a QC. Feynman proposed \cite{feynman82} that one of the most
important tasks that a QC could carry out is simulating quantum physics
in an efficient way, contrary to the inefficiency showed by classical 
computers. The Abrahams-Lloyd scheme can serve for implementing differential
equations in a QC \cite{differential}.

\section{The Schr\"odinger Equation and the Grover Algorithm}
\label{sec:3}

The initial steps pursued by Schr\"odinger to formulate his evolution equation
for the wave function of a quantum mechanical system bears little similarity 
with the final outcome he arrived at. 
In fact, he first came up with a scalar relativistic wave
equation, later known as the Klein-Gordon equation. The proccess of creation
of a new theory usually is not a straight line but rather a 
long a winding road,
whose final results are not as polished and neat as they look at the end.
Likewise, the final formulation of Grover algorithm does not show
the initial steps that his author followed. In fact, Grover wandered 
around the discretized time-dependent Schr\"odinger equation \cite{grover3} 
until he arrived to his beutiful algorithm in its final form 
\cite{grover2}, \cite{grover3}, \cite{groverfamily}.

The Grover algorithm is a quantum searching algorithm that outperforms any
classical searching algorithm. The problem of searching an item in a 
disordered list or database with $M$ items 
is known to be one of the most fundamental
tasks that any computer is oftenly doing. Classically, no algorithm can do
better than a brute force algorithm \cite{rmp}. 
The number of queries is order $O(M)$.
Remarkly enough, the quantum search algorithm can do better with the 
help of the quantum parallelism and constructive interference of amplitudes
in the evolution of a QC described in Sect.~\ref{sec:1}. In the quantum
case, the number of queries is reduced to order $O(\sqrt{M})$, which is a 
major improvement.

The basic steps that led Grover to the quantum search algorithm starting
from the discretized time-dependent Schr\"odinger equation are the
following:

\begin{description}

\item[1/] Discretize the time-dependent Schr\"odinger equation on a
space-time lattice with lattice spacing $a$ and time steps of $\tau$ duration
as described in Sect.~\ref{sec:2}, using an asymmetric difference scheme
(\ref{b5}), (\ref{b6}). This evolution is approximately unitary up to
order $O(\epsilon^2)+O(\tau^2)$.
\end{description}
It is interesting to notice that Grover does not resort to an exactly
reversible implementation of the evolution equation as in a 
symmetric difference scheme (\ref{b7}).
He employs a different pathway to achieve an exact unitary transformation
since he needs to stick to a first order difference equation in the time
variable.

\begin{description}

\item[2/] Introduce Diffusion $D$ and Rotation $R$ operators. This is 
a decomposition of the matrix evolution equation (\ref{b6}) as follows
\begin{equation}
\vec{\Psi}_{n+1}:= D R \vec{\Psi}_{n},
\label{c1}
\end{equation}
\begin{equation}
D := 
\left(\begin{array}{cccccc}
1 - 2{\rm i}\epsilon & 
{\rm i}\epsilon & 
0 & 
0 &
\ldots & 
{\rm i}\epsilon \\
{\rm i}\epsilon & 
1 - 2{\rm i}\epsilon &
{\rm i}\epsilon & 
0 &
\ldots & 
0 \\
\vdots &
\vdots &
\vdots &
\vdots &
\ddots &
\vdots \\
{\rm i}\epsilon & 
0 &
0 &
0 &
\ldots &
1 - 2{\rm i}\epsilon
\end{array} \right),
\label{c2}
\end{equation}
\begin{equation}
R := 
\left(\begin{array}{cccc}
{\rm e}^{-{\rm i}V_{1}\tau} & 
0 & 
\ldots & 
0 \\
0 & 
{\rm e}^{-{\rm i}V_{2}\tau} &
\ldots & 
0 \\
\vdots &
\vdots &
\ddots &
\vdots \\
0 & 
0 &
\ldots &
{\rm e}^{-{\rm i}V_{M}\tau}
\end{array} \right).
\label{c3}
\end{equation}
We have assumed that the potential is time independent. This decomposition
is correct within the approximation of order $O(\epsilon^2)+O(\tau^2)$.
Notice also that while $R$ is exactly unitary, $D$ is only within the
approximation $O(\epsilon^2)$.
\end{description}
The motivation for doing this decomposition is the key idea behind the
construction of the searching algorithm: the matrix $D$ may be interpreted
as a diffusion transformation (Markov process) with imaginary transition
probabilities ${\rm i}\epsilon$, and the matrix $R$ is a phase rotation at 
every site of
the lattice $m=1,2,\ldots,M$ induced by the application of the external
potential $V_{m}$. 

In the searching process, at time step $n=0$ the system is initially 
in an equally probable superposition of states from the database
to be searched for
\begin{equation}
\Psi_{m,0} := \frac{1}{\sqrt{M}}, \; \; m=1,2,\ldots,M.
\label{c3b}
\end{equation}
The goal is to find a marked state or item that is represented by a marked
component of the wave function $\Psi_{m_0,N}$, after a number $N$ of
iterations that represent the queries of the disordered database.
This can be achieved taming these transformations $D,R$ in order to
drive the system towards a final wave function that has the marked 
component highly amplified with respect to the rest of components.

 The diffusion process represented by $D$ affects
only to nearest-neighbour components of the wave function. This 
can also be viewed as the hopping of a single particle along the lattice
in the presence of a potential. A classical analogy is useful: we may
think that this particle will roll down towards the sites of lower 
potential like in a potential well. This is the amplification effect
needed in the quantum search.

\begin{description}

\item[3/] Global diffusion and localized rotation: it is advantageous
to extend the local diffusion of eq. (\ref{b2}) globally, as follows
\begin{equation}
D_{\rm L} := 
\left(\begin{array}{ccccc}
1 - (M-1){\rm i}\epsilon & 
{\rm i}\epsilon & 
{\rm i}\epsilon & 
\ldots & 
{\rm i}\epsilon \\
{\rm i}\epsilon & 
1 - (M-1) {\rm i}\epsilon &
{\rm i}\epsilon & 
\ldots & 
{\rm i}\epsilon \\
\vdots &
\vdots &
\vdots &
\ddots &
\vdots \\
{\rm i}\epsilon & 
{\rm i}\epsilon &
{\rm i}\epsilon &
\ldots &
1 - (M-1) {\rm i}\epsilon
\end{array} \right).
\label{c4a}
\end{equation}
Now, the particle may hop to any site of the lattice. This is an
example of long-range hopping. The sum of every column in (\ref{c4a}) is 1.

At the same time, we tune the form of the potential well to be localized
at the marked state as follows
\begin{equation}
R_{\rm L}(v) := 
\left(\begin{array}{ccccc}
1 & 
0 & 
0 &
\ldots & 
0 \\
0 & 
{\rm e}^{-{\rm i}v\tau} &
0 &
\ldots & 
0\\
0 & 
0 &
1 &
\ldots & 
0 \\
\vdots &
\vdots &
\vdots &
\ddots &
\vdots \\
0 & 
0 &
0 &
\ldots &
1
\end{array} \right),
\label{c4b}
\end{equation}
where $v$ is the value of the potential at that site. When it is negative
it means attraction. The marked state is unknown, but we can selectively
rotate its phase by an amount given by $v$.
\end{description}

With a global diffusion operation (\ref{c4a}) there is no net hopping 
between unmarked sites since they have the same amplitude and the
hopping in both directions cancels out. The hopping occurs towards the
marked site from the rest of all states. This is the mechanism of 
amplitude amplification in the search algorithm. This amplification
is maximum for a value of the rotation parameter $v=\frac{\pi}{2}$.
After this rotation and at time step $n$, 
let the amplitudes of the marked and unmarked states
be given by
\begin{equation}
\begin{array}{l}
\tilde{\Psi}_{m_0,n}  =  \frac{{\rm i}C_n}{\sqrt{M}},\; \; 
\tilde{\Psi}_{m\neq m_0,n}  =  \frac{c_n}{\sqrt{M}}.\\
\frac{C_n^2}{M} + c_n^2  =  1.
\end{array}
\label{c4c}
\end{equation}
Initially, $C_0:=1=:c_0$ (\ref{c3b}). After the diffusion transformation,
they become
\begin{equation}
\begin{array}{ccl}
\Psi_{m_0,n} & = & \frac{1}{\sqrt{M}}[{\rm i}C_n+({\rm i}c_n+C_n) (M-1)
\epsilon],\\
\Psi_{m\neq m_0,n} & = & \frac{1}{\sqrt{M}}[c_n-({\rm i}c_n+C_n) 
\epsilon].
\end{array}
\label{c4d}
\end{equation}
This means that the marked state changes its magnitude an amount
of the order $O(\sqrt{M}\epsilon)$, since by normalization 
(\ref{c4c}) $\frac{1}{2}<c_n<1$, assuming $C_n<\sqrt{\frac{M}{2}}$. 
We wish to make $\epsilon$ as large as possible in order to get the
highest magnitude change in the marked state. However, we recall that
the diffusion operation (\ref{c4a}) is unitary assuming that 
$\epsilon \ll \frac{1}{M}$. 

As the change of the marked state is order $O(\sqrt{M}\epsilon)$, if we
take $\epsilon = O(\frac{1}{M})$, then this change becomes 
$O(M^{-\frac{1}{2}})$. Thus, after the application of the diffusion
transformation (\ref{c4a}) a number of time steps $N=O(\sqrt{M})$,
the amplification of the marked state will become of the order of 1.

\begin{description}

\item[4/] Exactly unitary diffusion operation. Let us get rid off the
infinitesimal character of the diffusion (\ref{c4a}) introducing a
finite $D_{\rm F}$ transformation
\begin{equation}
D_{\rm F}(x,y) := 
\left(\begin{array}{cccccc}
x & 
y & 
y & 
y &
\ldots & 
y \\
y & 
x &
y & 
y &
\ldots & 
y \\
\vdots &
\vdots &
\vdots &
\vdots &
\ddots &
\vdots \\
y & 
y &
y &
y &
\ldots &
x
\end{array} \right),
\label{c5}
\end{equation}
where $x,y$ are complex amplitudes satisfying the unitary constraints
\begin{equation}
|x|^2 + (M-1) |y|^2 := 1, \; \; (xy^{\ast} + x^{\ast}y) + (M-2)|y|^2 :=0.
\label{c6}
\end{equation}
The infinitesimal transformation (\ref{c4a})  corresponds to the
particular case $x=1 - (M-1) {\rm i}\epsilon$, $
y={\rm i}\epsilon$. Solving (\ref{c6}) we find that the highest
value of $y$ can be obtained on real solutions as
\begin{equation}
x = -1 + \frac{2}{M}:=x_M, \; \; y = \frac{2}{M}:=y_M.
\label{c7}
\end{equation}
Then, the best diffusion operation is $D_{\rm F}(x_M,y_M)$ which
corresponds to a hopping amplitude as high as $\frac{2}{M}$.

\end{description}
This is the pathway followed by Grover in order to implement an 
exactly reversible evolution while retaining the first order character
of the iteration process.

\begin{description}

\item[5/] The quantum sarching algorithm: iterate a number $N$ of time
steps the basic operations (\ref{c5}) and (\ref{c4b}), 
starting from the uniformily distributed state (\ref{c3b}).
When the diffusion operation is exactly unitary (\ref{c5}), (\ref{c6}),
the maximum hopping into the marked site occurs for a rotation of
$v=\pi$ (\ref{c4b}), instead of $v=\frac{\pi}{2}$ that is the case when the
diffusion is infinitesimal (\ref{c4a}). Thus,
\begin{equation}
\vec{\Psi}_N = \left[ D_{\rm F}(x_M,y_M) R_{\rm L}(\pi)\right]^N \vec{\Psi}_0.
\label{c7b}
\end{equation}
This is an example of Trotter-Suzuki transformation \cite{trotter}, 
\cite{suzuki}. The counting of the number of iterations --queries of
the database-- is the same as in iv/ but with the application of
(\ref{c5}), (\ref{c6}) to the state
\begin{equation}
\begin{array}{l}
\tilde{\Psi}_{m_0,n}  =  \frac{-C_n}{\sqrt{M}},\; \; 
\tilde{\Psi}_{m\neq m_0,n}  =  \frac{c_n}{\sqrt{M}}.\\
\frac{C_n^2}{M} + c_n^2  =  1.
\end{array}
\label{c8}
\end{equation}
The result is,
\begin{equation}
\begin{array}{ccl}
\Psi_{m_0,n} & = & \frac{1}{\sqrt{M}}[C_n+\frac{2}{M}((M-1)c_n+C_n)],\\
\Psi_{m\neq m_0,n} & = & \frac{1}{\sqrt{M}}[-c_n+\frac{2}{M}(Mc_n-C_n)].
\end{array}
\label{c9}
\end{equation}
In the limit $M\rightarrow \infty$, 
the increase in the amplitude of the marked state is
of order $O(\frac{c_n}{\sqrt{M}})$, and the amplitude of the unmarked
states remains approximately the same. As $c_n$ remains order $O(1)$
during the evolution, the net amplification is order $O(M^{-\frac{1}{2}})$.
Therefore, we need a number of iterations (\ref{c7b}) of the order
$N=O(\sqrt{M})$ to make the probability of success for finding the 
searched item close to 1. 
\end{description}

The origin of the higher efficiency of the quantum search algorithm
over the classical algorithms is thus a neat combination of quantum
paralellism and constructive interference of amplitudes.
In addition, the Grover algorithm (\ref{c7b}) is an example of the 
evolution of a quantum system like in Sect.~\ref{sec:1}, 
with $\vec{\Psi}_N \rightarrow \Psi_{\rm f}$ and 
$\vec{\Psi}_0 \rightarrow \Psi_{\rm i}$,
in order to implement a useful task in Quantum Information Theory.

\section{Conclusions}
\label{sec:4}
It has been two decades now since the BB84 quantum cryptography
protocol was presented \cite{bb84} and one decade ago that Shor algorithm
was introduced \cite{shor}. I has been specially during this last decade
that Quantum Information Theory has emerged an matured as a solid and
complete discipline. QIT has part of its roots in the fundamentals of
Quantum Mechanics, and thus it has become a part of Theoretical Physics 
by now. It has ramifications in other disciplines like Information Theory,
Computer Science etc. which guarantees that findings in QIT may have
applications to other fields beyond. 
Specially rewarding is the fact that QIT is
a discipline in which there are experiments testing its theoretical findings.
All these aspects have made QIT a very exciting field so far and we 
expect that it will continue to be so in the future.

{\em Dedicated to Prof. Alberto Galindo Tixaire on occasion of his 
$70^{\rm th}$ birthday}.

\noindent {\em Acknowledgments}. This work is partially supported by the
DGES under contract BFM2003-05316-C02-01.
\appendix

\section{Numerical Stability of the Exactly Reversible Schr\"odinger Equation}
\label{sec:apA}

In the analysis of the finite difference equations obtained from the
time-dependent Schr\"odinger equation (\ref{b4}), whe have focused on
the exact reversibility on the lattice as the main issue.

Another very important issue is whether the difference equations are
well-posed: a solution of the difference equation must be a reasonable 
approximation to the solution of the differential equation. This is the
notion of {\em convergence}: in the limit of both the lattice spacing 
$a\rightarrow 0$ and the time step $\tau \rightarrow 0$, 
the solution of the difference equation must converge to the solution
of the differential equation. The convergence is measured with a
${\rm L}^2$-norm:
\begin{equation}
||\vec{\Psi}_n||_a^2 := \sum_m |\Psi_{m,n}|^2 a.
\label{apA1}
\end{equation}

A related issue is the {\em consistency} of the difference equation. It means
that the difference equation itself must converge to the differential equation
pointwise at each grid point as $a,\tau \rightarrow 0$. Consistency implies
that a smooth solution of the differential equation is an approximate solution 
of the difference equation.  While a direct proof of convergence
is a rather difficult problem, the proof of consistency is simpler.
For the exactly reversible Schr\"odinger difference equation (\ref{b7}), 
consistency is a direct consequence of the Taylor expansions 
(\ref{b2}), (\ref{b3}) and it is checked directly.

Consistency is a necessary condition for convergence, but it is not sufficient.
Additional properties must be required for a consistent difference equation
to be convergent. 
This is where the notion of {\em numerical stability} enters 
\cite{motter}, \cite{strikwerda}.
The essence of stability is that the solutions of the difference equation
must be bounded is some sense, in order to prevent the finite solution
to explode in the continuum limit. More precisely, a difference equation
is stable when the solution at a time step $n$ is bounded by
\begin{equation}
||\vec{\Psi}_n||_a^2 \leq C_T \sum_{n'=0}^{N_0} ||\vec{\Psi}_{n'}||_a^2,
\label{apA2}
\end{equation}
with $C_T$ some constant, and $N_0$ some integer that depends on the
order of the difference equation.
The meaning of this bound is that as time goes by, the growth of the solution  
is limited. This concept of stability is closely related
to the concept of well-posedness of initial value problems for partial 
differential equations:
\begin{equation}
\int_{-\infty}^{+\infty} |\Psi(x,t)|^2 dx \leq 
C_T \int_{-\infty}^{+\infty} |\Psi(0,t)|^2 dx, \; \; t\in[0,T].
\label{apA3}
\end{equation}

The fundamental theorem in the theory of finite difference equations is 
the Lax-Richtmyer equivalence theorem: a consistent finite difference
equation for a partial differential equation, for which the initial value
problem is well-posed, is convergent if and only if it is stable.

This theorem is a very powerful characterization of convergence since 
the determination of consistency and stability is much easier.
Let us apply the Lax-Richtmyer theorem to the exactly reversible 
Schr\"odinger difference equation (\ref{b7}) with vanishing external
potential. 
We already know that the equation is consistent from the
Taylor expansions (\ref{b2}), (\ref{b3}). Let us focus on stability.
In this particular case, we have translational invariance
and we also assume an infinite spatial lattice. 
This allows us to Fourier transform
to momentum space the solution $\Psi_{m,n}$ in coordinate space,
\begin{equation}
\hat{\Psi}_{n}(k) := \frac{1}{\sqrt{2\pi}} \sum_{m=-\infty}^{+\infty} 
{\rm e}^{-{\rm i}mak} \Psi_{m,n} a, \; \; k \in [-\frac{\pi}{a},\frac{\pi}{a}].
\label{apA4}
\end{equation}
Using the inversion formula
\begin{equation}
\Psi_{m,n} = \frac{1}{\sqrt{2\pi}} \int_{-\frac{\pi}{a}}^{+\frac{\pi}{a}} 
{\rm e}^{{\rm i}mak} \hat{\Psi}_{n}(k)dk, 
\label{apA5}
\end{equation}
we obtain the recursion relation satisfied by the Fourier components
\begin{equation}
\begin{array}{ccl}
\hat{\Psi}_{n+1}(k)& = &\hat{\Psi}_{n-1}(k) 
-2{\rm i}f_{\epsilon}(ka) \hat{\Psi}_{n}(k),\\
f_{\epsilon}(ka) & := & 4\epsilon \sin^2 \frac{ka}{2}.
\end{array}
\label{apA6}
\end{equation}
This is a simpler recursion relation since it is only on the time-step label
$n$.
The Parseval relation 
\begin{equation}
||\hat{\vec{\Psi}}_n||^2_a = \int_{-\frac{\pi}{a}}^{+\frac{\pi}{a}} 
|\hat{\Psi}_n(k)|^2 dk = \sum_{-\infty}^{+\infty} |\Psi_{m,n}|^2 a = 
||\vec{\Psi}_n||^2_a, 
\label{apA7}
\end{equation}
guarantees that the stability property is preserved
under the Fourier transform. There are several ways for solving (\ref{apA6}).
Let us introduce a generating function 
\begin{equation}
\hat{G}_k(u) := \sum_{n=0}^{\infty} \hat{\Psi}_n(k) u^n, 
\label{apA8}
\end{equation}
defined on the complex plane $u\in C$. Knowing $\hat{G}_k(u)$, 
we can recover the 
Fourier components through the Cauchy theorem,
\begin{equation}
\hat{\Psi}_n(k) = \frac{1}{2\pi {\rm i}} \oint_{\Gamma_0} 
\hat{G}_k(u) u^{-n-1} du,
\label{apA9}
\end{equation}
where $\Gamma_0$ is a contour encircling the origin counterclockwise.
Using the recursion relation we can find the generating function. It is
given by
\begin{equation}
\hat{G}_k(u) = 
\frac{[1+2{\rm i}f_{\epsilon}(ka) u]\hat{\Psi}_0(k) + u \hat{\Psi}_1(k)}
{1+2{\rm i}f_{\epsilon}(ka) u - u^2},
\label{apA10}
\end{equation}
where $\hat{\Psi}_0(k), \hat{\Psi}_1(k)$ are the Fourier transform of the
initial data (\ref{apA4}). To find the solution via (\ref{apA9}) 
it is convenient
to change variables $v:=\frac{1}{u}$,
\begin{equation}
\hat{\Psi}_n(k) = \frac{1}{2\pi {\rm i}} \oint_{\Gamma_{\infty}} 
\hat{G}_k(\frac{1}{v}) v^{n+1} \frac{dv}{v^2},
\label{apA11}
\end{equation}
where $\Gamma_{\infty}$ is a contour encircling the point of infinty and
enclosing all the poles of the generating function. Using Cauchy's residue's
theorem, we can obtain the general solution as
\begin{equation}
\begin{array}{ccl}
\hat{\Psi}_n(k) & = & \frac{1}{2\pi {\rm i}} \oint_{\Gamma_{\infty}} 
\hat{H}_k(v) v^{n} dv = \sum_{\rm poles} {\rm Res} [\hat{H}_k(v) v^{n}],\\
\hat{H}_k(v) & := & 
\frac{[v+2{\rm i}f_{\epsilon}(ka)]\hat{\Psi}_0(k) + \hat{\Psi}_1(k)}
{v^2+2{\rm i}f_{\epsilon}(ka) v - 1}.
\end{array}
\label{apA12}
\end{equation}
This solution depends on the roots of the equation
\begin{equation}
\begin{array}{l}
v^2+2{\rm i}f_{\epsilon}(ka) v - 1 = 0,\\
v_{\pm} = -{\rm i}f_{\epsilon}(ka) \pm \sqrt{1-f^2_{\epsilon}(ka)}.
\end{array}
\label{apA13}
\end{equation}
When the roots are distinct $v_{+}\neq v_{-}$, 
we have simple poles in (\ref{apA12}) and
the solution has the following form
\begin{equation}
\hat{\Psi}_n(k) = \frac{1}{2\sqrt{1-f^2_{\epsilon}(ka)}}
\sum_{s=\pm} s
\left[(v_{s}+2{\rm i}f_{\epsilon}(ka))\hat{\Psi}_0(k)+\hat{\Psi}_1(k)\right]
v_{s}^n. 
\label{apA14}
\end{equation}
The roots are equal $v_{+}:=v_0=: v_{-}$, when $f_{\epsilon}(ka)=\pm 1$
and then $v_0=\mp {\rm i}$. Thus,
we have a double pole in (\ref{apA12}) and
the solution takes the form
\begin{equation}
\hat{\Psi}_n(k) = \left[ (n-1) \hat{\Psi}_0(k)  \mp {\rm i}
n \hat{\Psi}_1(k) \right] {\rm e}^{\mp {\rm i}(n-2)\pi}.
\label{apA15}
\end{equation}

The stability analysis can be done with the help of the explicit solutions.
To simplify, let us assume that the limit $a,\tau \rightarrow 0$ is taken
with $\epsilon$ kept fixed \cite{strikwerda}.
When there are two different roots, the solution (\ref{apA14}) is bounded when
\begin{equation}
|v_{\pm}(ka)| \leq 1,
\label{apA16}
\end{equation}
and this determines the stability. If $4|\epsilon|>1$, then there exists
values of $ka$ for which $|v_{\pm}(ka)|>1$ and becomes unstable. 
It is when $4|\epsilon|\leq1$ that we can guarantee condition 
(\ref{apA16}) $\forall ka$,
since then $|v_{+}(ka)| = |v_{-}(ka)| =1$.

When there is a single root, the solution (\ref{apA15}) is not bounded due
to the linear growth with $n$. This instability occurs at 
$4|\epsilon| \sin^2\frac{ka}{2}=1.$ Since from the previous case we already
know that $4|\epsilon|$ can be at most 1, removing this possibility we
avoid the case of solution (\ref{apA15}). Therefore, the condition for
numerical stability is $4|\epsilon|<1$. In this case, the solution 
(\ref{apA14}) can be further simplified as
\begin{equation}
\hat{\Psi}_n(k) = \frac{1}{2\cos\theta_{\epsilon}(ka)}
\sum_{s=\pm} 
\left[{\rm e}^{{\rm i}s\theta_{\epsilon}(ka)}\hat{\Psi}_0(k)+
s \hat{\Psi}_1(k)\right]
s^{n}{\rm e}^{{\rm i}(1-s)n\theta_{\epsilon}(ka)}, 
\label{apA17}
\end{equation}
where we have introduced the following parametrization
\begin{equation}
f_{\epsilon}(ka) := \sin\theta_{\epsilon}(ka), \; \; 4|\epsilon|<1.
\label{apA18}
\end{equation}

Let us point out that the stability analysis carried out here can be
repeated for the exactly reversible Schr\"odinger equation in terms
of real and imaginary parts (\ref{b14}), (\ref{b15}). 
In this case we would need to introduce
a couple of generating functions, one for each component.


\begin{thebibliography}{99.}
\bibitem{functionevaluation} At first sight this splitting of registers
looks awkward but otherwise, we run into problems since the
direct function evaluation is doom to failure. Imagine we implement
the function evaluation as
\begin{equation}
U_f|x_1,x_2,\ldots,x_m\rangle = |f(x_1,x_2,\ldots,x_m)\rangle,
\label{bibi1}
\end{equation}
and $f$ is not a one-to-one application: there exists 
$(x_1,\ldots,x_m)\neq (y_1,\ldots,y_m)$ such that 
$f(x_1,\ldots,x_m)=f(y_1,\ldots,y_m)$. Choose the states
as orthonormal $\langle (x_1,\ldots,x_m)|(y_1,\ldots,y_m)\rangle=0$.
Then, the implementation (\ref{bibi1}) implies that the transformed
states are not orthogonal: 
$\langle f(x_1,\ldots,x_m)|f(y_1,\ldots,y_m)\rangle=1$.
This violates unitarity. In fact, we know that an orthonormal basis
of states is transformed onto another orthonormal basis of states
under a unitary transformation. Whence the necessity for splitting
registers (\ref{a6}),(\ref{a7}).
\bibitem{reversibility}

Irreversibility implies heat dissipation according to the second law
of thermodynamics: information cannot be lost and every erased or lost
bit of information is shed as a tiny puff of heat.
This shedding of information is happening permanently at each step of the fast 
and long chains of computations. Intermediate results in these calculations
are discarded since they are apparently useless and the computer needs
to reset the memory registers. This erasure produces heat that must be
removed from the machine to prevent from malfunctioning. The nature of this
heating is fundamental, intrinsic to information erasure.

There are other sources of heating in a computer like the dragging
of electrons in the currents inside the wires due to resistance. 
In actual computers this heat dissipation due to resistance is larger
by far than the heating due to information erasure. There is still a lot
of room for reducing resistance heating, for instance, introducing 
superconducting electron currents. When this would be achieved, then we
face the heating due to dumping information. Landauer first quantified
\cite{landauer61} in energy
this loss per bit as
\begin{equation}
\Delta E = k_{\rm B}T\ln 2.
\label{b0}
\end{equation}
This is an irreducible amount of heat that must unavoidably be shed when
a computer clears its registers by erasing a bit of information. 
The less information  systems erase, the less power they use.
Bennett
found \cite{bennett73} a way to overcome this heating introducing the 
concept of a reversible
computer that is universal as a Turing machine. Lecerf had introduced 
\cite{lecerf} this
idea of a reversible Turing machine earlier in the sixties.
In the Bennett machine, information is never throw out, thus causing no heat.
The problem then becomes one of a machine with an infinitely large memory.
To continue working, the computer needs somehow to be reset and here is where 
reversibility comes in: 
rewind the machine till the initial state without
erasing any single bit. These computers operate with reversible algorithms
that do not require data to be erased.
Bennett designed a memory-saving method in which a computer would 
carry out a few steps of the calculation, copy the result and rewind.
This is at the expense of processing time.

There has been recent interest in reversible classical computers
for reducing a system's heat level while allowing chip performance to scale.
The first reversible computer is a chip called Flattop \cite{flattop} 
developed at the
MIT's Reversible Computing research group. It is based on the reversible 
and adiabatic transfer of charge in digital circuits
implemented in standard silicon. 
Flattop implements the Billiard Ball Cellular Automaton \cite{billiardball}, 
a univeral and reversible model of computation introduced by Fredkin.
It is not a practical architecture for performing 
arbitrary computations, but it is regarded as a proof-of-concept reversible
machine showing that universal reversible computers may be accessible in
the near future.
\end{thebibliography}
\end{document}